\documentclass[twocolumn,aps,showpacs,floatfix,amsmath,amssymb,superscriptaddress,prb]{revtex4}

\usepackage{graphicx}
\usepackage{bm}
\usepackage{dsfont}
\usepackage{mathtools}
\usepackage[colorlinks=true,citecolor=blue,urlcolor=blue]{hyperref}

\newcommand{\bc}{\begin{center}}
\newcommand{\ec}{\end{center}}
\newcommand{\be}{\begin{equation}}
\newcommand{\ee}{\end{equation}}
\newcommand{\ba}{\begin{array}}
\newcommand{\ea}{\end{array}}
\newcommand{\beq}{\begin{eqnarray}}
\newcommand{\eeq}{\end{eqnarray}}
\newcommand{\ket}[1]{\left| {#1}\right\rangle}

\newcommand{\e}[1]{\langle {#1}\rangle}
\newcommand{\abs}[1]{\left| {#1}\right|}

\begin{document}

\title{Griffiths singularities in the random quantum Ising antiferromagnet: a tree tensor network renormalization group study}
\author{Yu-Ping Lin}
\affiliation{Department of Physics, National Taiwan University, Taipei 10617, Taiwan}

\author{Ying-Jer Kao }
\affiliation{Department of Physics, National Taiwan University, Taipei 10617, Taiwan}
\affiliation{National Center of Theoretical Sciences, National Tsing Hua University, Hsinchu 300, Taiwan}

\author{Pochung Chen}
\affiliation{Department of Physics, National Tsing Hua University, Hsinchu 300, Taiwan}

\author{Yu-Cheng Lin}
\email{yc.lin@nccu.edu.tw}
\affiliation{Graduate Institute of Applied Physics, National Chengchi University, Taipei, Taiwan}

\begin{abstract} 
The antiferromagnetic Ising chain in both transverse and longitudinal magnetic
fields is one of the paradigmatic models of a quantum phase transition.  The
antiferromagnetic system exhibits a zero-temperature critical line separating
an antiferromagnetic phase and a paramagnetic phase; the critical line connects
an integrable quantum critical point at zero longitudinal field and a classical
first-order transition point at zero transverse field.  Using a strong-disorder
renormalization group method formulated as a tree tensor network, we study the
zero-temperature phase of the quantum Ising chain with bond randomness.  We
introduce a new matrix product operator representation of high-order moments,
which provides an efficient and accurate tool for determining quantum phase
transitions via the Binder cumulant of the order parameter. Our results
demonstrate an infinite-randomness quantum critical point in zero longitudinal
field accompanied by pronounced quantum Griffiths singularities, arising from
rare ordered regions with anomalously slow fluctuations inside the paramagnetic
phase. The strong Griffiths effects are signaled by a large dynamical exponent
$z>1$, which characterizes a power-law density of low-energy states of the
localized rare regions and becomes infinite at the quantum critical point.
Upon application of a longitudinal field, the quantum phase transition between
the paramagnetic phase and the antiferromagnetic phase is completely destroyed.
Furthermore, quantum Griffiths effects are suppressed, showing $z<1$, when the
dynamics of the rare regions is hampered by the longitudinal field.
\end{abstract}

\date{\today}

\maketitle

\section{Introduction}
\label{intro}

The interplay between quantum fluctuations and randomness often leads to
drastic disorder effects, resulting in some exotic phenomena at and near
zero-temperature phase transitions.  For example, there exists one class of
disordered systems where strong Griffiths
singularities,~\cite{Griffiths,McCoy,McCoy_Wu} which arise from the formation
of rare strongly coupled region, lead to the divergence of the dynamical
exponent at quantum critical points.  The random transverse-field Ising chain
is the most prominent example for such spectacular phenomena and its quantum
critical point is, in the renormalization-group language, described as an {\it
infinite-randomness fixed point}, where the system appears more and more
disordered under coarse graining.~\cite{Fisher_PRL,Fisher_Ising,Motrunich}
There are many other one-dimensional random quantum systems that also exhibit
infinite-randomness phases, such as the random singlet phases of SU(2)$_k$
anyonic chains for all $k\ge 2$ (including the case $k\to \infty$ which
corresponds to the random spin-1/2 Heisenberg chain)~\cite{Fisher_AF,Bonesteel}
and the random J-Q spin-1/2 chains with multi-spin couplings.~\cite{SDRG_JQ} It
is noteworthy that the infinite-randomness fixed points can control more than
just one dimension; in particular, the quantum phase transitions in higher
dimensions with Ising (Z$_2$) symmetry have been numerically demonstrated to be
of infinite-randomness type.~\cite{QMC_2D,Lin_2D,Istvan_2D}  Experimentally,
Griffiths singularities have been observed in various quantum phases with
quenched disorder.~\cite{Neto98,Aeppli07}  The signature of an
infinite-randomness phase with the diverging dynamical critical exponent has
also been detected in recent experimental studies on random Heisenberg
chains~\cite{RS_exp,RS_exp1} and the superconductor-metal quantum phase
transition of a two-dimensional superconducting system.~\cite{SC_exp}

The Griffiths singularity was first discussed nearly fifty years ago in the
context of the randomly dilute Ising ferromagnet;~\cite{Griffiths} it arises
because there are in general rare but arbitrarily large spatial regions which
contain no vacancies and therefore show local magnetic order even when the
system is globally in the paramagnetic phase.  Such  locally ordered regions
produce non-analyticity in the free energy at zero field, known as the
Griffiths singularity, in the paramagnetic phase below the critical temperature
of the clean model.  The same rare region effect can be generalized to
non-dilute random-bond systems.  The Griffiths singularity of the free energy
in the context of a thermal transition is in general an essential
singularity,~\cite{Wortis,Harris_G,Shankar} and is thus too weak to be observed
experimentally.  By contrast, effects of rare regions on dynamical properties
are more promising to detect in experiments or numerical
simulations;~\cite{Palmer} in particular, in a quantum system, where both
dynamics and statics are inextricably coupled, the rare regions are strongly
correlated along the time-like direction, which enhances the Griffiths
phenomena.~\cite{Thill,Fisher_PRL,Fisher_Ising,Young_QCP1996,Young_GS1996,Vojta_Rev}

Below we use a simple quantum spin chain with binary disorder distribution to
explain the origin of the quantum Griffiths singularity. 
Generalizations to other disorder distributions and higher dimensions are
straightforward.~\cite{Thill,Vojta_Rev}  
We consider here a transverse-field
Ising model with a constant transverse field $\Gamma=1$ perpendicular to the
Ising axis and with random bonds ($J$) drawn from the binary distribution:
$\pi(J)=p\delta(J-\lambda)+(1-p)\delta(J-\lambda^{-1})$, where $\lambda\gg 1$
and $p<1$, corresponding to a quantum paramagnet.~\cite{RW,Frechet}  
The probability of finding a strongly coupled region of $n$ strong bonds
($J=\lambda$) decreases exponentially with the volume as $P(n)\approx p^n$. 
The energy gap of such a rare region is also exponentially small in its volume
$\varepsilon \approx 1/\lambda^n$, corresponding to an exponentially large
relaxation time.  
Thus, by change of variables we obtain a power-law
density of states for the small energy gaps: 
\be
     \rho(\varepsilon) \sim (\ln \lambda)^{-1} \varepsilon^\omega\,,
    \label{eq:P_e}
\ee  
with $\omega=-\ln(p)/\ln(\lambda)-1$. In a finite chain of length $L$, 
from the integrated density of states $N_\varepsilon=\int_0^\varepsilon d\varepsilon' \rho(\varepsilon')\sim L^{-1}$
we arrive at the relation between the length scale and the energy scale:
\be
    \varepsilon \sim L^{-1/(\omega+1)} \sim L^{-z}
    \label{eq:e_L_z}
\ee 
where we have introduced the dynamical exponent $z=(\omega+1)^{-1}$, which varies continuously with the disorder parameter $p$
and becomes larger and larger as the quantum critical point ($p=1$) is approached.
The power-law density of low energy excitations 
is responsible for singular low-energy behavior of various observables away from the critical point; for example the
average local susceptibility varies with the temperature as 
\be
       \chi^{\text{loc}}(T) \sim T^{-1+1/z}\,,
\ee
which diverges at zero temperature $T\to 0$ if $z>1$, even though the system is in the paramagnetic phase.

Many quantitative and exact asymptotic results for the scaling behavior in the
Griffiths phase and in particular at the random quantum critical point
can be obtained by a powerful renormalization group method, called the {\it
strong-disorder renormalization group} (SDRG), which was first introduced by Ma
{\it et al.}~\cite{Ma,Dasgupta} and extended by Fisher~\cite{Fisher_Ising}
and others.~\cite{SDRG_rev}
In the SDRG picture, the rare regions that give rise to
quantum Griffiths singularities are strongly correlated sites connected with
effective interactions generated during the action of the renormalization.
Analytical SDRG results for the transverse-field Ising chain show that with any
bond randomness the length dependence of a typical energy gap at the quantum
critical point has, in contrast to the power law  in  Eq.~(\ref{eq:e_L_z}),  an
exponential form:~\cite{Fisher_PRL,Fisher_Ising}
\be
 %-\ln(\epsilon) \sim L^{1/2}\,, 
 %\varepsilon \sim \exp(-\text{const. } L^{1/2})\,,%\quad c=\text{const.} 
 \varepsilon_\text{typ} \sim \exp(-c L^{1/2})\,,\quad c=\text{const.}
 \label{eq:activated} 
\ee 
corresponding to activated dynamical scaling, and $z\to\infty$;
on the other hand, the average gap scales as~\cite{Fisher_Young}
\be
 %-\ln(\overline{\epsilon}) \sim L^{1/3}\,,
 %\overline{\varepsilon} \sim \exp(-\text{const. } L^{1/3})\,,%\quad c'=\text{const.}
 \overline{\varepsilon} \sim \exp(-c'  L^{1/3})\,,\quad c'=\text{const.}
 \label{eq:avg_gap}
\ee
where the overbar denotes an average over disorder realizations. 
This unconventional dynamical scaling along with the distinction between average and typical (i.e. most probable) values
characterize an infinite-randomness fixed point, where energy gaps become
broadly distributed even on the logarithmic scale.  
Away from the infinite-randomness critical point in the Griffiths phase, the distribution of
low energy excitations and the dynamical exponent $z$ can also be obtained
within the SDRG framework.~\cite{Igloi}
In the paramagnetic phase, where the rare regions are localized, the distribution of their smallest energy gaps is
identified as a Fr\'echet-type distribution.~\cite{Fisher_Young,Frechet}
Interestingly, many analytical SDRG results for one-dimensional systems,
including the relations given in Eqs.~(\ref{eq:activated}) and
(\ref{eq:avg_gap}), can also be obtained from a variety of statistical
models.~\cite{Monthus,McKenzie,Monthus_RW,Z}

Here we investigate strong disorder effects on the antiferromagnetic Ising spin
chain in transverse and longitudinal magnetic fields using a generalization of
SDRG represented as a tree tensor network.~\cite{TSDRG}
We introduce a novel matrix product operator representation of high-order moments of the staggered
magnetization and evaluate the Binder cumulant to identify the quantum critical
points. 
In the absence of disorder, the quantum spin chain is
antiferromagnetic in weak fields at zero temperature, and paramagnetic in
strong fields; the two phases are separated by a critical line ending at a
classical multicritical point where the transverse field vanishes.  
In the presence of bond randomness, in addition to the infinite-randomness critical
point and Griffiths regions in zero longitudinal field, we find that the phase
transition is destroyed and quantum Griffiths effects are weakened upon a longitudinal
field is applied.

The paper is organized  as follows. In Sec.~\ref{sec:model} we
define the model and summarize some known properties of the zero-temperature
phases. 
In Sec.~\ref{sec:MPO} we describe the matrix product operator
representations used for our calculations; in particular, we 
propose a new representation of high-order moments.
In Sec.~\ref{sec:TSDRG} we extend the technique
used in Ref.~\onlinecite{TSDRG} and introduce the tensor network SDRG for the quantum
Ising chain with bond randomness plus on-site disorder, and provide the
results.  
We conclude in Sec.~\ref{sec:discussion} with a summary and
discussion.

\section{The model}
\label{sec:model}
 We study  the  antiferromagnetic Ising chain in the presence of both longitudinal and transverse magnetic fields, described by the Hamiltonian: %for a chain with $L$ spins:
\be
     %H=\sum_{i=1}^{L} J_i \sigma_i^z \sigma^z_{i+1}-\sum_{i=1}^L h_i^x \sigma_i^x -\sum_{i=1}^L h_i^z\sigma_i^z\,,
     H=\sum_{i} J_i \sigma_i^z \sigma^z_{i+1}-\sum_{i} \Gamma_i \sigma_i^x -\sum_{i} h_i\sigma_i^z\,,
     \label{eq:H}
\ee  
where $\sigma_i^z\,,\sigma_i^x$ are Pauli spin operators, $h_i$ ($\Gamma_i$) is the longitudinal (transverse) field applied to site $i$, 
and $J_i>0$ is a nearest-neighbor antiferromagnetic interaction which favors staggered magnetic ordering along the $z$ axis.

The Hamiltonian of the quantum spin chain defined in Eq.~(\ref{eq:H})
can be mapped to a classical two-dimensional Ising model using the path integral
formalism, where the extra dimension is regarded as imaginary time~\cite{Sachdev} and has extent $1/T$.
In this mapping the quantum model effectively corresponds to
the classical model described by
\be
     H_{\text{cl}}=\sum_{i,n} K_i s_{i,n} s_{i+1,n} - \sum_{i,n} K'_i s_{i,n} s_{i,n+1} -\sum_{i,n} h'_i s_{i,n}\,,
     \label{eq:mapping}
\ee
where $s_{i,n}=\pm 1$ and $n$ is the index in the imaginary time direction, running
from 0 to $1/(T\Delta\tau)$ with $\Delta\tau$ being the width of a time slice;
the corresponding couplings and fields are given by
\beq
        & K_i &=J_i \Delta\tau \\
        & e^{-2K'_i} &= \Gamma_i \Delta\tau \\
        & h'_i &=h_i \Delta\tau\,.
\eeq
To suppress the errors from the Suzuki-Trotter expansion~\cite{Suzuki,Trotter} used in this formalism, one requires $\Delta\tau \ll 1$, implying strong ferromagnetic couplings
$K'_i\gg 1$ along the imaginary time direction.
As a result, the classical model is strongly anisotropic, 
equivalent to ferromagnetic Ising chains weakly antiferromagnetically coupled with each other.

\subsection{Clean system}
We consider first the model with a homogeneous (i.e. site-independent) coupling, $J_i=1,\,\forall i$, and 
homogeneous applied fields $\Gamma_i=\Gamma,\,h_i=h,\,\forall i$. 
With $h=0$, the model is exactly solvable and is equivalent to the ferromagnetic quantum Ising chain through a gauge transformation.
At absolute zero temperature, the model with zero $h$ exhibits two phases: 
an antiferromagnetic ordered phase for $\Gamma<1$ and a
paramagnetic phase for $\Gamma >1$.
The phase transition at $\Gamma=\Gamma_c=1$ between these two phases is characterized by a vanishing
energy gap and a diverging correlation length. 
As $\Gamma$ approaches the critical value, the lowest energy gap vanishes as~\cite{Sachdev} 
\be
      %\epsilon \sim \abs{h^{x,c}-h^x}\,, 
      \varepsilon \sim \abs{1-\Gamma}\,,
\ee
and the correlation length diverges as
\be
      % \xi \sim \abs{h^x-h^{x,c}}^{-1}\,,
      \xi \sim \abs{1-\Gamma}^{-1}\,,
\ee
implying the dynamical exponent $z=1$ and the correlation length critical exponent $\nu=1$. 
This zero-temperature phase transition belongs to the two-dimensional Ising  universality class.

%%%%%%%%%%%%%%%%% FIG1 %%%%%%%%%%%%%%%%%%%%%%%
\begin{figure}
\centerline{\includegraphics[width=8.6cm, clip]{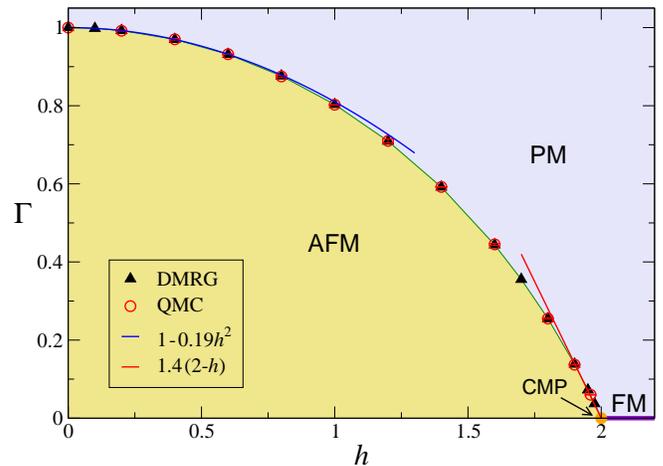}}
%\vskip-2mm
\caption{(Color online) The zero-temperature phase diagram of the clean antiferromagnetic Ising chain with $J=1$
in transverse ($\Gamma$) and longitudinal ($h$) magnetic fields.
The critical line, obtained by DMRG and QMC calculations,
separates a quantum antiferromagnetic (AFM) phase in weak fields and a quantum paramagnetic (PM) phase
in strong fields. The classical multicritical point (CMP) at $(\Gamma=0,\,h=2)$ 
corresponds to a first-order transition between an AFM phase and a ferromagnetic (FM) phase
in the absence of quantum fluctuations.  
The critical line is well described by a quadratic relation: $\Gamma_c\approx 1-0.19 h^2$ in the vicinity
of the point $(\Gamma=1,\,h=0)$, while it is linear, $\Gamma_c\approx 1.4(2-h)$,
near the classical multicritical point.
}
\label{fig:clean_phase}
%\vskip-3mm
\end{figure}
%%%%%%%%%%%%%%%%%%%%%%%%%%%%%%%%%

With finite value of $h$, the model is distinct from the ferromagnetic system.
In the ferromagnetic case the quantum phase transition occurs only in zero
longitudinal field $h=0$, while in the antiferromagnetic model the phase
transition remains at $h \neq 0$, yet the longitudinal field can further
destabilize antiferromagnetic order. 
As a result, the critical value $\Gamma_{c}$ at $T=0$ decreases as the strength of the longitudinal field
increases and the critical points terminate at $\Gamma=0,\,h=2$, where a
classical first-order transition takes place, corresponding to a multicritical
point. 
The quantum critical line in the $\Gamma-h$ plane except for the two
end points at $(\Gamma,\,h)=(1,\,0)$ and $(0,\,2)$ is not exactly solvable;
nevertheless, it falls into the two-dimensional Ising universality
class.~\cite{AF_DMRG}

For clarity, we present here the zero-temperature phase diagram in Fig.~\ref{fig:clean_phase}, reproduced by
quantum Monte Carlo (QMC) and density matrix renormalization group (DMRG) calculations.
The QMC algorithm is an unbiased method
working in terms of the formalism given in Eq.~(\ref{eq:mapping}) in the limit
of  $\Delta \tau\to 0$.~\cite{Cluster}
Some details of the DMRG algorithm are given in Sec.~\ref{sec:MPO}.  
Perturbation theory  predicts that the critical line drops quadratically with increasing $h$
in the limit of $h\to 0$, and is linear in the vicinity of the multicritical point.~\cite{AF_DMRG}
Our data agree with this asymptotic behavior:
the critical points for $h\le 0.8$ are well described by a quadratic function
\be
   \Gamma_{c} \sim 1 -a h^2
   \label{eq:quadratic}
\ee
with $a\approx 0.19$;
near the multicritical point, the critical line drops linearly as
\be
   \Gamma_{c} \sim b(2-h)
   \label{eq:linear}
\ee
with an estimated prefactor $b\approx 1.4$.

The classical multicritical point at $(\Gamma=0,\,h=2)$ is a first-order
transition point separating an antiferromagnetic ground state for $h<2$ and a
ferromagnetic ground state for $h>2$. 
The ground state is extensively
degenerate at the critical field $h=2$, where in a staggered
$\uparrow\downarrow\uparrow\downarrow\cdots$ configuration any spin pointing in
the opposite direction to the field can be overturned without change of
energy.~\cite{Domb}

\subsection{Random-bond chain with $h=0$}
Consider now the model with quenched disorder in which the interactions $J_i$
and/or the transverse fields $\Gamma_i$ are independent random variables.
In the space-time lattice model described in Eq.~(\ref{eq:mapping}), the randomness
introduced into $J$ and $\Gamma$ produces inhomogeneities extending only in spatial coordinates
but it is perfectly correlated in the imaginary time direction.
The random-bond quantum model in zero longitudinal field is effectively 
a solvable two-dimensional anisotropic Ising model;~\cite{McCoy_Wu,Shankar}
it undergoes a $T=0$ phase transition when~\cite{Shankar}
\be
    \overline{\ln J}=\overline{\ln \Gamma}\,.
    \label{eq:duality}
\ee
The critical behavior
is governed by an exactly solvable infinite-randomness fixed point
within the framework of SDRG.
At this fixed point, in addition to the energy gap,
spin correlations $C(r)=(-1)^r\e{\sigma_i^z\sigma_{i+r}^z}$ are also very broadly distributed,~\cite{Fisher_Ising}
with the typical behavior
\be
    -\ln C_{\text{typ}}(r) \sim \sqrt{r}\,.
\ee
The average correlation function, which is experimentally detectable, is however dominated by rare pairs of widely separated spins
that have a correlation of order unity, much larger than the typical value. 
Thus, different from the typical correlations, the average correlations decay as a power of $r$ at the critical point,~\cite{Fisher_Ising}
\be
      \overline{C}(r)\sim \frac{1}{r^{2-\phi}}\,,
\ee
where
$ \phi=(1+\sqrt{5})/2
      %\phi=\frac{1+\sqrt{5}}{2}
$
is the golden mean. 
For an open chain of length $L$, 
one is often interested in the end-to-end correlation function ${C}_{1,L}=(-1)^{L-1}\e{\sigma_1^z\sigma_{L}^z}$,
which considers correlations between two end spins; the average of ${C}_{1,L}$ is
%the average end-to-end correlation $\overline{C}_{1,L}=(-1)^{L-1}\e{\sigma_1^z\sigma_{L}^z}$ is
also dominated by rare samples for which the two end spins are almost perfectly correlated with each other, 
and it decays algebraically as
\be
     \overline{C}_{1,L}\sim \frac{1}{L}
     \label{eq:C_1L}
\ee
at criticality.~\cite{Fisher_Young,RW}

For a random chain, the deviation from criticality is often parameterized by~\cite{Fisher_Ising}
\be
 \delta=\frac{\overline{\ln \Gamma}-\overline{\ln J}}{\text{var}(\ln h)+\text{var}(\ln J)}\,,    
 \label{eq:delta}
\ee
where $\text{var}(x)$ stands for the variance of the random variable $x$.
Slightly away from criticality $\delta\neq 0$, the distinction between the average and the typical correlations
leads to two correlation lengths.~\cite{Fisher_Ising,Fisher_Young}
In the disordered phase $\delta>0$, the average correlations, both $\overline{C}(r)$ and $\overline{C}_{1,L}$, 
decay exponentially with the true correlation length $\xi \sim 1/\delta^2$, implying the correlation length exponent $\nu=2$ 
for the random chain.
The typical correlations decay even faster and its associated correlation length, $\xi_{\text{typ}}$, varies as
$\xi_{\text{typ}}\sim 1/\delta$, defining another exponent $\nu_{\text{typ}}=1$. 

The dynamical exponent $z$ which characterizes the Griffiths singularity 
in the off-critical region varies with the distance $\delta$ from the critical point.
In fact, this nonuniversal exponent is the positive root of the equation:
\be
      \overline{\left(\frac{J}{\Gamma} \right)^{1/z}}=1\,,
      \label{eq:z_eq}
\ee
which is an exact expression in the entire Griffiths region.~\cite{Z,Igloi}
In the vicinity of  the critical point, the dynamical exponent to the leading order is given by~\cite{Fisher_Ising,Igloi}
\be
       z\approx \frac{1}{2\abs{\delta}},\qquad \abs{\delta} \ll 1\,.
\ee 

\subsection{Random-bond chain with $h\neq 0$}

In addition to the clean system and the random chain in the absence of
longitudinal fields, we also consider the random chain in a finite longitudinal
field, which to our knowledge has not been previously studied.

We will show in Sec.~\ref{sec:TSDRG} that with finite $h$ the quantum phase transition
between the paramagnetic ground state and the antiferromagnetic ground state is
completely smeared out.
  Also the pronounced Griffiths effects seen in the zero
$h$ regime are strongly reduced upon a longitudinal field is applied.

\section{Matrix product operator representations}
\label{sec:MPO}

We have studied the ground-state properties and the zero-temperature phase diagram of the quantum antiferromagnetic Ising chain using tensor networks.
All calculations in our study involving tensor networks were performed using the Uni10 library.~\cite{uni10}
The starting point of our calculations is to construct the matrix product operator (MPO) representations~\cite{Scholl_Rev,McCulloch}
of the Hamiltonian and observables.
The Hamiltonian described in Eq.~(\ref{eq:H}) for a chain of $L$ spins with periodic boundary conditions (PBC) can be written as an MPO 
with operator-valued matrices ${W}_{H_p}^{[i]}$ (where $H_p$ stands for the Hamiltonian with PBC) at sites $i$ in the bulk: 
\begin{equation}
\begin{split}
{W}_{H_p}^{[i]}  & =
\begin{pmatrix}
\mathds{1} & 0 & 0 & 0 \\
0 & \mathds{1} & 0 & 0 \\
\sigma^z_i & 0 & 0 & 0 \\
-\Gamma_i\sigma^x_i-h_i\sigma^z_i & 0 & J_i \sigma^z_i & \mathds{1}
\end{pmatrix}\,,\\
 &\qquad\qquad\qquad\qquad 2\leq i \leq L-1\,,
\end{split}
\end{equation}
and at the first and last sites: 
\begin{equation}
{W}_{H_p}^{[1]} =
\begin{pmatrix}
-\Gamma_1\sigma^x_1-h_1\sigma^z_1\; &
J_L\sigma^z_1\; &
J_1\sigma^z_1\; &
\mathds{1}
\end{pmatrix}\,,
\end{equation}

\begin{equation}
%\begin{split}
{W}_{H_p}^{[L]}  =
\begin{pmatrix}
\mathds{1} \\
\sigma^z_L \\
\sigma^z_L \\
-\Gamma_L\sigma^x_L-h_L\sigma^z_L
\end{pmatrix}\,.
%\,,\\
% &\qquad\qquad\qquad\qquad 2\leq i \leq L-1\,,
%\end{split}
\end{equation}
On an open chain, the Hamiltonian (denoted by $H_o$) is encoded by the following MPO tensors:
\begin{equation}
 \begin{split} 
{W}_{H_o}^{[i]} &=
\begin{pmatrix}
\mathds{1} & 0 & 0  \\
\sigma^z_i & 0 & 0 \\
-\Gamma_i\sigma^x_i-h_i\sigma^z_i & J_i\sigma^z_i & \mathds{1}
\end{pmatrix}\,,\\
 &\qquad\qquad\qquad\qquad 2\leq i \leq L-1\,,
\end{split}
\end{equation}
%and edge matrices:
\begin{equation}
{W}_{H_o}^{[1]} =
\begin{pmatrix}
-\Gamma_1\sigma^x_1-h_1\sigma^z_1\; &
J_1\sigma^z_1\; &
\mathds{1}
\end{pmatrix}\,,
\end{equation}

\begin{equation}
%\begin{split}
{W}_{H_o}^{[L]}  =
\begin{pmatrix}
\mathds{1} \\
\sigma^z_L \\
-\Gamma_L\sigma^x_L-h_L\sigma^z_L
\end{pmatrix}\,.
\end{equation}

%%%%%%%%%%%%%%%%% FIG2 %%%%%%%%%%%%%%%%%%%%%%
\begin{figure}
\centerline{\includegraphics[width=8.6cm, clip]{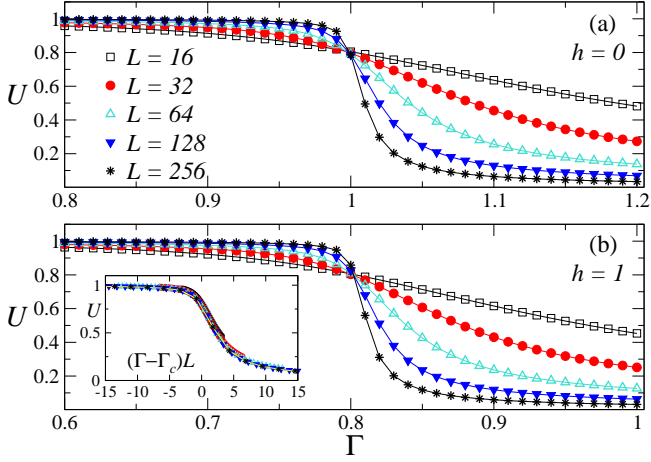}}
\caption{\label{fig:clean_U}(Color online)
The Binder cumulant $U$ of the order parameter of the clean
chain with $J=1$
in the absence of longitudinal fields (a) and in a finite longitudinal field
$h=1$ (b),  plotted against the strength of the transverse field. 
The crossing point of the curves for different chain lengths $L$
indicates the critical value $\Gamma_c$ of the transverse field for each case; we find $\Gamma_c=1$ for $h=0$ and $\Gamma_c=0.8$ for $h=1$.
The inset of (b) is a scaling plot
of the data for $h=1$, using the critical exponent $\nu=1$ of the 2D Ising
universality class.}
%\label{fig:clean_U}
\end{figure}
%%%%%%%%%%%%%%%%%%%%%%%%%%%%%%%%%%

The Binder cumulant,~\cite{Binder} which involves the second and the fourth moments of the order parameter, 
is a commonly used quantity in numerical studies for determining the critical point and critical exponents.
The calculation of the expectation value of a high-order moment in a tensor network state is in general not 
straightforward.~\cite{Wei}
Here we introduce an MPO representation for the high-order moments of an observable, which allows an efficient computation of the Binder ratio. 
 
Consider an $n$-th moment of some observable $O$ of the form
\be
   O^n=\left(\frac{1}{L} \sum_{i=1}^L O_i \right)^n\,,
  %Q^n=\left(\sum_i O_i \right)^n\,,
\ee
where $i$ runs over all the sites in the system and $O_i$ is the on-site operator at $i$.
Here we introduce a generic MPO form for $O^n$ with arbitrary $n$, 
\be
   O^n=\frac{1}{L^n} W_{O^n}^{[1]}W_{O^n}^{[2]}\cdots W_{O^n}^{[L]} \,,
\ee
where the edge matrices are given by 
\be
    W_{O^n}^{[1]}=
    \begin{pmatrix}
    O_1^n\; &
    \binom{n}{1}O_1^{n-1}\; &
    \binom{n}{2}O_1^{n-2}\; &
    \cdots & 
    \mathds{1}
    \end{pmatrix}\,,
\ee
and
\be
    W_{O^n}^{[L]}=\begin{pmatrix}
    \mathds{1} \\
     O_L \\
     O_L^2 \\
      \;\vdots \\
     O_L^n 
     \end{pmatrix}\,,
\ee
with the binomial coefficient $\binom{n}{k}$,
%$X^L=(0,\cdots,0,\mathds{1})$ and $X^R=(\mathds{1},0,\cdots,0)^T$, and
and bulk matrices 
$W_{O^n}^{[i]}$ at sites $i=2\cdots L-1$ read
\be
    W_{O^n}^{[i]}=
     \begin{pmatrix*}[l]
       \,\mathds{1} & & & \vspace{5pt} \\
	O_i & \;\mathds{1} & & & \vspace{5pt} \\
	O_i^2 & \binom{2}{1}O_i & \;\mathds{1} & & \\
	\;\vdots & & & \ddots & \\
	O_i^n & \binom{n}{1}O_i^{n-1} & \binom{n}{2}O_i^{n-2} & \cdots & \mathds{1} 
     \end{pmatrix*}\,.
     \label{eq:On}
\ee
%\textcolor{red}{This is a first explicit MPO construction of an $n$-th moment with bond dimension growing {\it linearly} with $n$.} 
Expressing an $n$-th moment in terms of this MPO, the bond dimension grows {\it linearly} with $n$. One
can directly use this explicit MPO construction for any power $n$ without using MPO compression to reduce the bond dimensions,
which is the advantage of this MPO representation over previously proposed constructions.~\cite{McCulloch,Generic_MPO}

Using the density matrix renormalization group (DMRG)
technique~\cite{White_DMRG} in terms of matrix product states~\cite{McCulloch,Scholl_Rev} and the MPO representations described above, we
determine the ground state of the chain with PBC in the absence of disorder.
For a fixed $h$, we find the critical value of $\Gamma$ by studying the Binder cumulant $U$ of the staggered magnetization,
defined as~\cite{Sandvik_Rev} 
\be
     U=\frac{1}{2}\left(3-\frac{\e{m_s^4}}{\e{m_s^2}^2} \right)\,,
     \label{eq:Binder}
\ee
where the staggered magnetization $m_s$ is given by
\be
  m_s=\frac{1}{L}\sum_{i=1}^L (-1)^i \sigma_i^z\,,
\ee
and $\e{\cdots}$ denotes the expectation value in the ground state.
The curves of the Binder cumulant as a function of $\Gamma$ at fixed $h$ for
different system sizes $L$ cross each other, as illustrated in
Fig.~\ref{fig:clean_U}.  
The critical values $\Gamma_c$ can then be extracted
from the crossings of the Binder cumulant.  
In the inset of Fig.~\ref{fig:clean_U}(b), we have rescaled the Binder cumulant
for $h=1$ horizontally, multiplying $(\Gamma-\Gamma_c)$ by $L^{1/\nu}$ with $\nu=1$,
to achieve data collapse; the same scaling form applies 
for other critical points plotted in the phase diagram in Fig.~\ref{fig:clean_phase} except for
the classical multicritical point at $h=2$, where no crossing point is found in 
the curves of the Binder cumulant $U(\Gamma)$ for different system sizes.
We have examined the accuracy of the DMRG results by comparing the locations of
the critical points in Fig.~\ref{fig:clean_phase} with those obtained by the
continuous-time quantum Monte Carlo (QMC) algorithm.
Excellent agreement can be observed between the results obtained from the two methods.

We conclude this section with the note that the MPO representation for
high-order  moments and cumulants introduced here circumvents tedious
calculations involving sums of correlators $O_{i_1}O_{i_2}\cdots O_{i_n}$.
This new MPO requires no additional truncation to represent an $n$-th moment
with a bond dimension linear in $n$, and provides an efficient  way to
determine the critical point with great accuracy, also for the disordered
system discussed in the next section.

\section{Tree tensor network strong disorder renormalization group} 
\label{sec:TSDRG}

The DMRG method can be applied to systems with quenched disorder, too.
However, a practical implementation of the DMRG for disordered systems is
computationally expensive because large bond dimensions and many DMRG sweeps
are often required to avoid getting stuck in local minima.~\cite{Hubbard1,Hubbard2} 

The Hamiltonian of the quantum spin model in Eq.~(\ref{eq:H}) with randomness
has an intrinsic separation in energy scales, which allows us to find its
ground state using the SDRG technique.  The basic strategy of the SDRG method,
introduced in Ref.~\onlinecite{Ma} and Ref.~\onlinecite{Fisher_Ising}, is to
find the largest term in the Hamiltonian successively and lock the spins
associated with this term into their local ground state.  For example, if the
largest term is a field, $\Gamma$ or $h$, on spin $i$, then that spin is frozen
in the $x$- or $z$-direction; if, on the other hand, the largest term in the
Hamiltonian is an interaction $J_{ij}$, the two spins associated with this term
are combined together into an antiferromagnetic cluster, in the state
$\ket{\uparrow_i\downarrow_j}$ or $\ket{\downarrow_i\uparrow_j}$.  The largest
term in the Hamiltonian is thus effectively eliminated and energy corrections
in terms of effective interactions or fields are generated using perturbation
theory.  This RG process yields an effective Hamiltonian with gradually fewer
degrees of freedom and lower energy scale.

Although the SDRG procedure is approximate, it yields asymptotically exact results
(up to logarithmic corrections~\cite{SDRG_JQ,XX_Log}) in the low-energy limit
for the scaling behavior of systems at quantum critical points governed by an
infinite-randomness fixed point, where the widths of the distributions of the
{\it logarithmic} effective couplings diverge.  
To justify the accuracy of this
approximate RG in general cases, one should carefully examine whether the
condition of validity of perturbation theory is satisfied, that is, whether the
largest energy scale to be decimated is much larger than those of the
neighboring terms (the perturbation).  
This condition is often not satisfied in the early stage of RG when the disorder distribution is not very broad.
One approach to refine the perturbative approximation is to include higher energy
states, rather than only the lowest multiplet,  at each RG
step;~\cite{Hikihara} this extended SDRG method was further
reformulated in terms of tree tensor networks in Ref.~\onlinecite{TSDRG} for the random-bond
Heisenberg chain.  
Below we demonstrate that the tree tensor network SDRG is applicable to 
the random quantum Ising chain also with site disorder.

\subsection{Numerical scheme}

%%%%%%%%%%%%%%%%% FIG3 %%%%%%%%%%%%%%%%%%%%%%
\begin{figure}
\begin{center}
\includegraphics[width=8.6cm, clip]{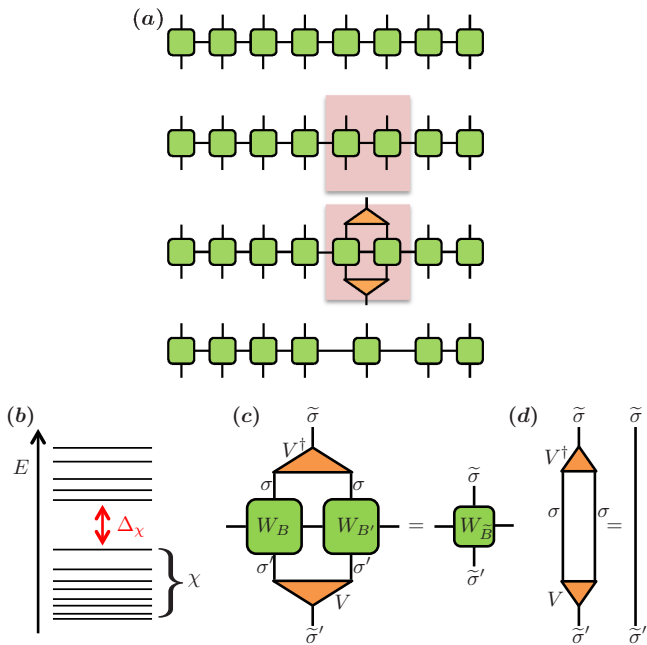}
\caption{\label{fig:RG}
(Color online) Schematic of the tensor network SDRG algorithm. (a) The system is partitioned into blocks using MPO formalism; the green boxes represent the $W$ tensors, the vertical lines are physical indices and the horizontal lines represent the bond indices. 
At each step of RG, the pair of nearest-neighbor blocks (marked with the pink-shaded square) 
with largest energy gap $\Delta_\chi$ is chosen for renormalization, where
$\Delta_\chi$, as sketched in (b),  is the gap between the highest energy of the $\chi$ lowest energy states that would be kept and the 
lowest-energy multiplet that would be discarded. The $\chi$ lowest energy eigenvectors of the two-block Hamiltonian are used to create an isometric tensor
$V$, represented by a three-leg triangle, which satisfies $V^\dagger V=\mathds{1}$ shown in subfigure (d). 
The subfigure (c) illustrates the truncation of the two blocks with the largest $\Delta_\chi$ into one block in terms of isometric tensors.    
}
\end{center}
\vskip-3mm
\end{figure}
%%%%%%%%%%%%%%%%%%%%%%%%%%%%%%%%%%

For clarity of  presentation, here we review the formulation of SDRG as a tree tensor network as proposed in Ref.~\onlinecite{TSDRG}.

A natural way to include the effect of excited states of local Hamiltonians
in the RG process is to partition the system into blocks.~\cite{Hikihara}
In each RG iteration, we compute the energy spectrum of each nearest-neighbor
two-block Hamiltonian and identify the energy gap $\Delta_\chi$, which is
measured as the difference between the highest energy of the $\chi$-lowest
energy states that would be kept and the higher multiplets that would be discarded.
We then combine the pair of blocks with the largest energy gap $\Delta_\chi$
into one block. The RG process is repeated
until the chain is described by one block.

%%%%%%%%%%%%%%%%% FIG4 %%%%%%%%%%%%%%%%%%%%%%
\begin{figure}
%\centerline{\includegraphics[width=8cm, clip]{Isometry.eps}\includegraphics[width=8cm, clip]{Isometry.eps}}
\begin{center}
\includegraphics[width=8.6cm, clip]{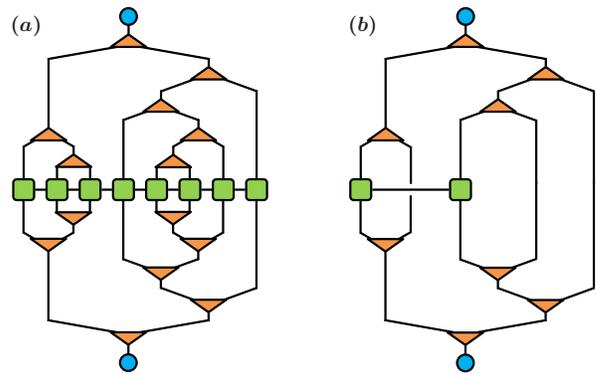}
\caption{\label{fig:TTN}
(Color online)(a) The full SDRG algorithm can be seen as a binary tree tensor network with an inhomogeneous structure. 
The circles represent the ground state eigenvector of the final block resulting from the SDRG procedure.
(b) An example of calculating the ground-state expectation value of some operator which acts only on two sites, such as
the two-point correlation function; only those tensors associated with the two sites that the operator acts on need to be considered.  
}
\end{center}
\end{figure}
%%%%%%%%%%%%%%%%%%%%%%%%%%%%%%%%%%

The SDRG procedure formulated as tree tensor networks can be described as follows~\cite{TSDRG} (see Fig.~\ref{fig:RG}):
\begin{itemize}
\item[1.] Decompose the chain into a set of $n$-site blocks and construct the block MPO tensors $W_B$.
\item[2.] Obtain the energy spectrum  of the two-block Hamiltonian for each pair of nearest-neighbor blocks.
\item[3.] Merge the pair of blocks (say $B$ and $B'$) with the largest energy gap $\Delta^{\text{max}}\equiv\max\{\Delta_\chi\}$ into a new block and contract the MPO tensors to form $W_{BB'}$.
\item[4.] Build a rank-3 isometry tensor $V$ with the $\chi$ lowest energy states of new block as column vectors; 
use $V$ and $V^\dagger$ to truncate $W_{BB'}$ to $W_{\tilde{B}}$. %See Fig.~\ref{x}
\item[5.] Repeat steps 2-4 until the system is represented by one single block.
\end{itemize}
Finally, we diagonalize the Hamiltonian $H_{B_f}$ of last one block resulting from the RG process to obtain the
eigenvalues, which represent the low-energy spectrum of original system. 
The ground-state expectation value of some observable can be obtained 
by contracting the MPO with the tree tensor network composed of 
the ground state of $H_{B_f}$ and the isometric tensors $V$ created during the RG.

\subsection{Results}

Using the tensor network SDRG we investigate the $T=0$ phase diagram of the random-bond antiferromagnetic Ising
chain in longitudinal and transverse magnetic fields, which to our knowledge has not been previously studied.
We use open boundary conditions and the following distributions of the random bonds and random 
transverse fields:
\be
  \begin{split}
    \pi_1(J) &=
    \begin{cases}
      1 & \hspace*{0.65cm}\text{for } 0<J\le 1\,,\\
      0 & \hspace*{0.65cm}\text{otherwise.}
    \end{cases} \\
    \pi_2(\Gamma) &= 
    \begin{cases}
      1/\Gamma_0 & \text{for } 0<\Gamma \le \Gamma_0\,,\\
      0 & \text{otherwise.} 
    \end{cases}
  \end{split} 
  \label{eq:J_distrib} 
\ee
The width, $\Gamma_0$, of the field distribution is therefore the single
control parameter in our simulations.
We have achieved convergence for system sizes up to $L=128$ 
using four-site blocks ($n=4$) and $\chi=50$. 
In addition, for each case more than 10000 independent disorder realizations (samples) 
were used to obtain the disorder averages with sufficiently small error bars.
In the following, we discuss the results for the case with
zero longitudinal field ($h=0$) and with nonzero longitudinal field ($h\neq 0$),
separately.

\subsubsection{Zero longitudinal field}
\label{sec:random_h0}
%%%%%%%%%%%%%%%%% FIG5 %%%%%%%%%%%%%%%%%%%%%%
\begin{figure}
\centerline{\includegraphics[width=8.6cm, clip]{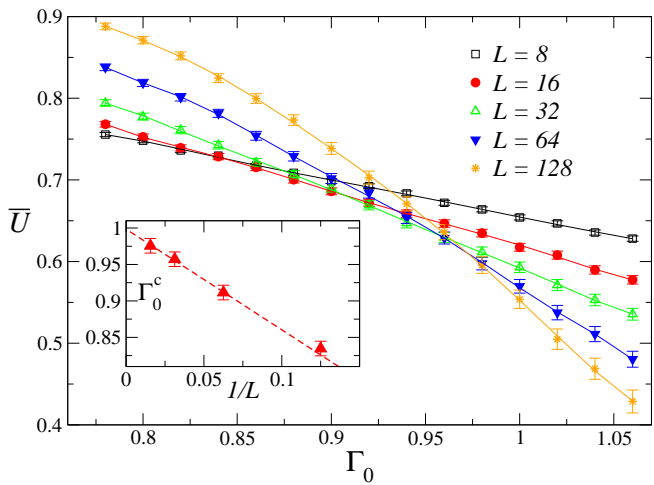}}
\caption{\label{fig:U_dis_z0}(Color online)
The disorder-averaged Binder cumulant $\overline{U}$ of the random
chain in the absence of longitudinal fields. The curves for
different sizes are graphed versus the cut-off $\Gamma_0$ of the random transverse fields.
The inset is a plot for extracting the location of the disordered
quantum critical point by using the crossings of $\overline{U}$ for
system sizes $(L,2L)$; from an extrapolation to $1/L\to 0$, we obtain
the critical value $\Gamma_0^c=1$.}
\end{figure}
%%%%%%%%%%%%%%%%%%%%%%%%%%%%%%%%%%

We begin first with the random model in the absence of a longitudinal field.
The location of the critical point for this case is determined by
the relation given in Eq.~(\ref{eq:duality}) or Eq.~(\ref{eq:delta}).
For the distributions of Eq.~(\ref{eq:J_distrib}), the critical point
occurs when $\Gamma_0=1$, which can be verified by using 
the Binder cumulant MPO described in Sec.~\ref{sec:MPO} for finite-size systems.
To obtain the disorder average of the Binder cumulant we first average
the expectation values of the moments over samples and then evaluate the ratio, i.e.
\be
     \overline{U}=\frac{1}{2}\left(3-\frac{\overline{\e{m_s^4}}}{\overline{\e{m_s^2}}^2} \right)\,;
     \label{eq:Binder_dis}
\ee 
in this way we reduce errors in evaluating the ratio.~\cite{Kawashima}
As shown in Fig.~\ref{fig:U_dis_z0}, the curves of the Binder cumulant for
different system sizes cross each other, although the crossings exhibit some
significant drift for the finite open chain considered here; the finite-size
scaling plot in the inset of this figure shows that an extrapolation of the
crossings  to $L\to \infty$ based on data sets for sizes $L$ and $2L$ indeed
gives the critical value $\Gamma_0^c=1$.  
We note that computing the Binder
cumulant using the conventional SDRG method does not appear to have a straightforward
implementation, which shows a clear advantage of the tensor network approach.

%%%%%%%%%%%%%%%%% FIG6 %%%%%%%%%%%%%%%%%%%%%%
\begin{figure}
\centerline{\includegraphics[width=8.6cm, clip]{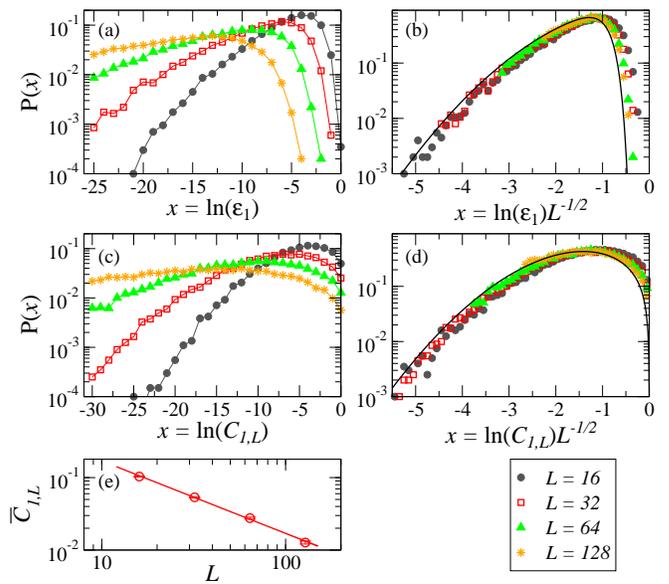}}
\caption{\label{fig:EC_z0_x1}(Color online)
Distributions of energy gaps (a) and end-to-end correlation functions (c)
at the critical point ($h=0,\,\Gamma_0=1$) for different sizes.
Subfigures (b) and (d) are scaling plots for the data in (a) and (c), respectively;
the solid lines in the scaling plots are the analytical results given in Ref.~\onlinecite{Fisher_Young}. 
Subfigure (e) shows the mean correlations; the solid red line has a slope of $-1$,
in agreement with the theoretical prediction Eq.~(\ref{eq:C_1L}).
%(a) The distribution of energy gaps at the critical point ($h=0,\,\Gamma_0=1$) for different sizes; the distribution was
%obtained from more than 20000 samples for each size.
%(b) Scaling plot of the gaps distribution;
%(c) The distribution of the end-to-end correlation function ${C}_{1,L}$ at the critical point;
%(d) A scaling plot of the ${C}_{1,L}$ distribution. 
%The solid lines in (c) and (d) are
%the analytical results given in Ref.~\cite{Fisher_Young}. 
%Subfigure (e) shows the mean correlations; the solid red line has a slope of $-1$,
%in agreement with the theoretical prediction Eq.~(\ref{eq:C_1L}).
}
\end{figure}
%%%%%%%%%%%%%%%%%%%%%%%%%%%%%%%%%%

The random quantum critical point with $h=0$ is an infinite-randomness fixed
point, where the dynamical exponent diverges $z\to \infty$, characterized by a
stretched exponential scaling form of energy gaps (see
Eq.~(\ref{eq:activated})).  
We have determined the energy gap from the
lowest-lying excitation, denoted by $\varepsilon_1$, of the final normalized
block.
  The distribution of energy gaps at the critical point and a scaling
plot using the scaling variable $x=\ln(\varepsilon_1)/\sqrt{L}$ are shown in
Fig.~\ref{fig:EC_z0_x1}(a) and Fig.~\ref{fig:EC_z0_x1}(b), respectively, where the scaling function agrees well with the
analytical expression given in Ref.~\onlinecite{Fisher_Young},
which has a limiting form as $P(x)\approx 2\pi^{-1/2}\exp(-x^2/4)$ for large $\abs{x}$
and approaches $P(x)\approx -2\pi x^{-3}\exp(-\pi^2/(4x^2))$ for small $\abs{x}$.

We have also calculated the end-to-end correlation function $C_{1,L}$. 
 Like the distribution of energy gaps, the distribution of the typical correlations
$C_{1,L}$ (shown in Fig.~\ref{fig:EC_z0_x1}(c)) at the critical point is extremely broad even on a logarithmic scale
and its width grows with increasing $L$.
%
%  Fig.~\ref{fig:EC_z0_x1}(c) and Fig.~\ref{fig:EC_z0_x1}(d) show the
%distribution and 
Fig.~\ref{fig:EC_z0_x1}(d) shows a scaling plot with the rescaled variable
$x=\ln(C_{1,L})/\sqrt{L}$; here the scaling function is also found to be in
good agreement with the analytic prediction:
$P(x)=-x/2\exp(-x^2/4)$.~\cite{Fisher_Young}
While the typical correlations
decay exponentially with the distance $L$,  the average value of $C_{1,L}$,
shown in Fig.\ref{fig:EC_z0_x1}(e), falls off according to
$1/L$, \cite{Fisher_Young} and is dominated by rare events in which the two end
spins have strong correlations of order unity.  

%%%%%%%%%%%%%%%%% FIG7 %%%%%%%%%%%%%%%%%%%%%%
\begin{figure}
%\centerline{\includegraphics[width=8.6cm, clip]{fig8.eps}}
\centerline{\includegraphics[width=8.6cm, clip]{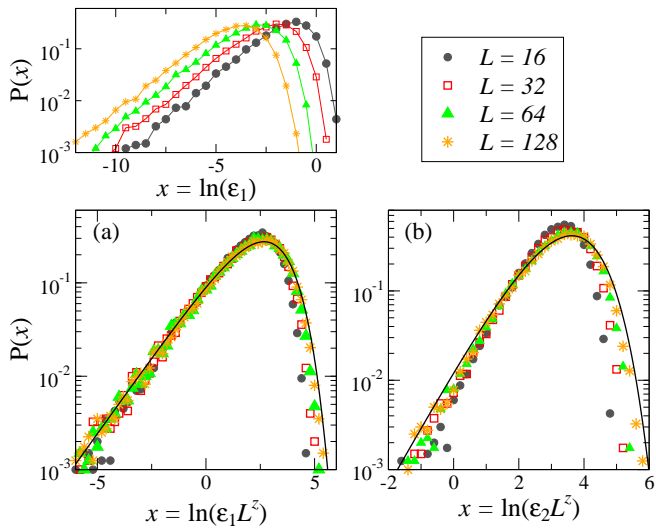}}
\caption{\label{fig:E_z0_x3}(Color online)
Finite size scaling of the distributions of the first excitation energy $\varepsilon_1$ (a) and the second excitation energy $\varepsilon_2$ (b) 
in the Griffiths phase at $h=0$ and $\Gamma_0=3$. 
The solid lines are plots of the Fr\'echet distribution in Eq.~(\ref{eq:frechet}) with the dynamical exponent $z=1.335$, which is the solution of Eq.~(\ref{eq:z_box}), and a fitting parameter $\alpha=0.07$. The upper figure shows the unscaled distribution of $\ln(\varepsilon_1)$. 
}
\end{figure}
%%%%%%%%%%%%%%%%%%%%%%%%%%%%%%%%%%

Away from the quantum critical point in the paramagnetic phase where $\Gamma_0>1$,
the dynamical exponent $z$ for the whole region is also known from analytical results~\cite{Z,Igloi}
and is given by the root of Eq.~(\ref{eq:z_eq});
for the distribution of Eq.~(\ref{eq:J_distrib}), the dynamical exponent is then
determined by the equation
\be
    z\ln\left(1-z^{-2} \right)=-\ln \Gamma_0\,.
    \label{eq:z_box}
\ee  
Furthermore, deep in the paramagnetic phase $\delta\gg 1$ where the
low-energy excitations are rare, the distributions of $k$th low-lying
excitation energies $\varepsilon_k$ for a finite chain of size $L$ have the
following scaling form~\cite{Young_GS1996,Igloi_3z,Frechet}
\be
    P(\varepsilon_k)=L^z \tilde{P}_k(u_k)\,,
   \label{eq:e_scaling} 
\ee
where the scaling function $\tilde{P}_k(u_k)$, based on extreme-value statistics,~\cite{Extreme} is described by
the Fr\'echet distribution given by~\cite{Frechet}
\be
   \tilde{P}_k(u_k)=\frac{1}{z} u_k^{k/z-1} \exp\bigl(-u_k^{1/z}\bigr)\,,
   \label{eq:frechet}
\ee
in terms of the variable
\be
      u_k=\alpha \varepsilon_k L^z\,,
\ee
with a nonuniversal constant $\alpha$.
In Fig.~\ref{fig:E_z0_x3} the distributions of the first and second excitation energy of the final RG block are 
scaled according to Eq.~(\ref{eq:e_scaling}) with the dynamical exponent $z\approx 1.3354$ obtained from Eq.~(\ref{eq:z_box}).
The large dynamical exponent $z>1$ is a signature of the quantum Griffiths singularity. 
In fact, all solutions of Eq.~(\ref{eq:z_box})
for $\Gamma_0>1$ give $z>1$, the value of which increases with decreasing distance to the quantum critical point $\Gamma_0=1$.

\subsubsection{Nonzero longitudinal field}
\label{sec:random_h}

Now we turn to the random chain in the presence of a longitudinal field and
with the rectangular distributions of Eq.~(\ref{eq:J_distrib}) for the bonds
$J_i$ and transverse fields $\Gamma_i$. 
Without loss of generality, we consider
the case where the strength of the longitudinal field is a site-independent
constant $h>0$.

%%%%%%%%%%%%%%%%% FIG8 %%%%%%%%%%%%%%%%%%%%%%
\begin{figure}
\centerline{\includegraphics[width=8cm, clip]{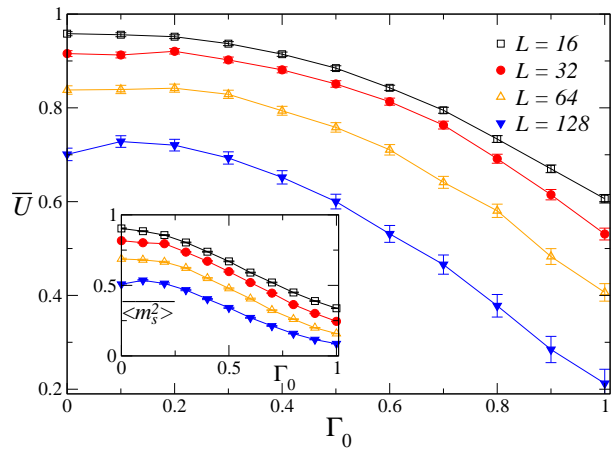}}
%\centerline{\includegraphics[width=8cm, clip]{Bin_dis_z002.eps}}
\caption{\label{fig:U_dis_z002}(Color online)
The disorder-averaged Binder cumulant $\overline{U}$ of the random
chain in a longitudinal field of strength $h=0.02$, graphed versus
the cut-off $\Gamma_0$ of the random transverse fields.
The inset: the disorder average of the squared staggered magnetization as a function of $\Gamma_0$.
}
\end{figure}
%%%%%%%%%%%%%%%%%%%%%%%%%%%%%%%%%%

Figure~\ref{fig:U_dis_z002} shows the disorder-averaged Binder cumulant,
$\overline{U}$, of the staggered magnetization as a function of the cut-off of
the transverse field, ranged from $\Gamma_0=0$ to 1, at a small longitudinal
field $h=0.02$. 
 The curves for different system sizes have no crossing, giving
evidence that the staggered magnetization does not show a phase transition.
Furthermore, all values of $\overline{U}$ as well as the squared
antiferromagnetic order parameter $\overline{\e{m_s^2}}$ (shown in the inset)
decrease with increasing $L$, indicating vanishing antiferromagnetic order in
this region where the longitudinal field is finite but only slightly away from
zero. 
 We recall that the chain in the classical limit $\Gamma_0=0$ is
antiferromagnetic only in the state which consists of alternating spins
pointing in opposite directions; in the chain with bond randomness ($J_i\in
(0,1]$ in our case), a finite magnetic field will break up the staggered order
of the spins if some bonds are weaker than the field, which explains the
destruction of the antiferromagnetic order in the whole regime where $h>0$,
including the ''quantum'' phase with finite transverse fields. 
%A schematic phase diagram of the random chain is shown in Fig.~\ref{fig:diagram_dis}. 

%%%%%%%%%%%%%%%%% FIG9 %%%%%%%%%%%%%%%%%%%%%%
\begin{figure}
%\centerline{\includegraphics[width=8.6cm, clip]{fig10.eps}}
\centerline{\includegraphics[width=8.6cm, clip]{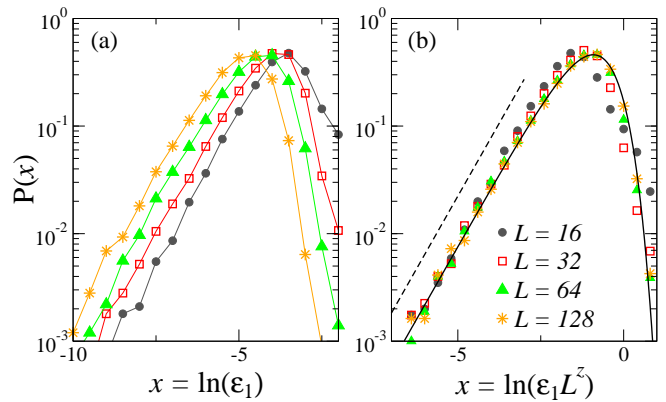}}
\caption{\label{fig:E_z002_x1}(Color online)
(a) The distribution of the log of the first excitation energy $\varepsilon_1$ at $\Gamma_0=1\,\,h=0.02$, 
in the vicinity of infinite-randomness
quantum critical point. (b)  Scaling plot of the data in (a) using the scaling variable $x=\ln(\varepsilon_1L^z)$. 
The black dashed line indicates the slope of low-energy tail, which gives $z\simeq 0.8$; the black solid line is a plot of the Fr\'echet distribution given in Eq.~(\ref{eq:frechet}). 
%The inset shows the average excitation energy $\overline{\varepsilon}_1$ versus the chain length $L$, which gives
%$z\simeq 0.7$. 
}
\end{figure}
%%%%%%%%%%%%%%%%%%%%%%%%%%%%%%%%%%

Next, we analyze disorder effects in the finite-$h$ paramagnetic phase near the
infinite-randomness fixed point at $\Gamma_0=1$. 
 In Fig.~\ref{fig:E_z002_x1}
the distribution of first energy gaps at $\Gamma_0=1,\,h=0.02$ for different
system sizes is shown.  Here the distribution of $\ln(\varepsilon_1)$ looks
similar to the distribution in the Griffiths phase at $\Gamma_0>1,\,h=0$ (see
Fig.~\ref{fig:E_z0_x3}); the low-energy tail of the distribution for different
sizes is just shifted horizontally relative to each other and shows power-law
behavior. 
 According to Eq.~(\ref{eq:frechet}), the power of the low-energy
tail of $P(\ln(\varepsilon_1))$ is given by $1/z$ (here $k=1$).  
We estimate $z\simeq 0.8$ from the slope of the tails in Fig.~\ref{fig:E_z002_x1}, and find
good agreement between the scaling function, in terms of the scaling variable
$\varepsilon_1 L^z$, and the Fr\'echet distribution described in
Eq.~(\ref{eq:frechet}).  
The power-law tail and the Fr\'echet distribution of
lowest gaps show that these low-energy excitations are localized in this phase.
The small dynamical exponent $z<1$ does not lead to divergence  
of the linear susceptibility, although it can produce divergence in 
the non-linear susceptibility at $T=0$ when $z>1/3$.~\cite{Young_GS1996, Igloi_3z}
In comparison with the situation in the zero-$h$ paramagnetic phase, the 
dynamical exponent here in the vicinity of the quantum critical point is much smaller.
The strongly reduced Griffiths effects can be understood as follows.
A rare region in the equivalent classical system
corresponds to a quasi-one-dimensional ferromagnetic Ising model along
the time-like direction. 
An applied longitudinal field will
align the spins in the rod-like rare region and suppress fluctuations of the 
order parameter, %so that the rare region behaves essentially classically, 
so that the rare region becomes essentially static,
leading to a suppression of quantum Griffiths effects.

%%%%%%%%%%%%%%%%% FIG10 %%%%%%%%%%%%%%%%%%%%%%
\begin{figure}
%\centerline{\includegraphics[width=8.6cm, clip]{fig10.eps}}
\centerline{\includegraphics[width=8.6cm, clip]{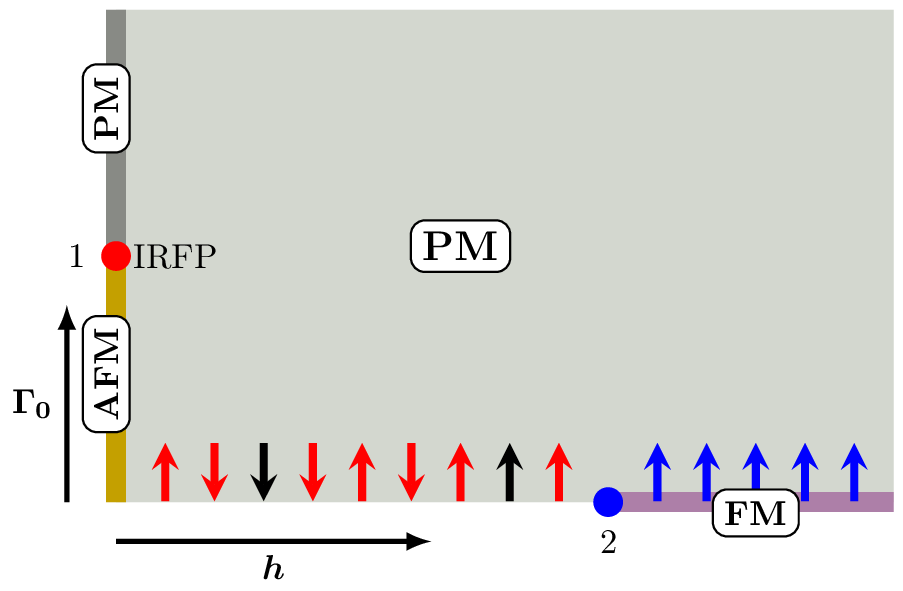}}
\caption{\label{fig:diagram_dis}(Color online)
Sketch of the $h-\Gamma_0$ phase diagram of the random-bond Ising antiferromagnetic chain with $J_i\in (0,1]$.
The ground state exhibits a quantum phase transition governed by an infinite-randomness fixed point (IRFP)
between an antiferromagnetic (AFM) phase at zero-$h$ and a paramagnetic (PM) phase with 
pronounced quantum Griffiths effects (the dark gray strip along the $\Gamma_0$-axis), 
resulting from rare ordered regions with anomalously slow fluctuations.
In the finite-$h$ regime, the AFM order is destroyed even in the classical limit $\Gamma_0\to 0$,
and also the quantum Griffiths effects are weakened.
}
\end{figure}
%%%%%%%%%%%%%%%%%%%%%%%%%%%%%%%%%%

\section{Discussion}
\label{sec:discussion}

Using the DMRG and the SDRG algorithms based on matrix product operators, we
have explored the zero-temperature phases of the antiferromagnetic Ising chain
in longitudinal ($h$) and transverse ($\Gamma$) magnetic fields.  We have
introduced a new matrix product operator technique for calculating expectation
values of high-order moments, which allows us to compute the Binder cumulant of
the order parameter efficiently and locate the quantum critical points
with high accuracy, both in the clean system and in the system with disorder.

%The clean system at zero temperature exhibits an antiferromagnetic phase in
%weak fields, and a paramagnetic phase in strong fields. 
%These two phases in the
%$\Gamma-h$ plane are separated by a critical line, starting from the quantum
%critical point at zero $h$ and terminating at the classical
%antiferromagnetic-ferromagnetic first-order transition point at zero $\Gamma$.
Quenched disorder and rare regions of local order give rise to drastic effects
on the zero-temperature phases.  
The random quantum critical point triggered by
the transverse field in the absence of the longitudinal component is of
unconventional type governed by an infinite-randomness fixed point; also
pronounced Griffiths effects exist in the zero-$h$ paramagnetic phase,
characterized by a large dynamical exponent ($z>1$) in the entire region.  

We have unambiguously found that the aforementioned infinite-randomness quantum
critical point in the disordered system are
destroyed by the longitudinal component of the applied field. 
A schematic phase diagram of the random chain is shown in Fig.~\ref{fig:diagram_dis}.
The chain with
bond randomness exhibits no antiferromagnetic order even in the classical limit
(with zero transverse component of the field) when the applied field disturbs
the perfect staggered spin configuration by flipping any spin over;  this
explains the destruction of antiferromagnetic order and quantum criticality in
the random chain with finite $h$ that we consider.  
Furthermore, the
longitudinal field hampers the quantum tunneling of rare ordered regions by
pinning, %leading to classical behavior of the rare regions and the suppression of
leading to the suppression of
quantum Griffiths behavior in the finite $h$ regime. Similar phenomena also
occur in metallic Ising magnets.~\cite{Millis_PRL,Millis_PRB,Vojta_rounding}

The model we consider in this paper is interesting in many aspects.  First, the
model in the presence of a longitudinal field is nonintegrable even in the
clean case, offering a playground to explore how dynamics depends on the
integrability and the nonintegrability.~\cite{clean_dyn, clean_dyn_f} Second,
the model with bond-randomness has unconventional ground-state properties as
discussed in this paper, and due to its strong disorder effects the model
provides a theoretical paradigm for the problem of
localization.~\cite{dyn2012,Altman2014,RSRG-X,Imbrie} 
Furthermore, the model is
experimentally realizable using ultracold spinless bosonic atoms confined in a tilted
optical lattice with a magnetic field gradient along one direction, where each Ising spin is encoded in the motional degree
of freedom of a single atom.~\cite{cold_atom} In such an ultracold atomic
system, the longitudinal and transverse field are represented by the lattice
tilt and tunneling of the atom, respectively; the spin-spin interaction arises from
a nearest-neighbor constraint in the site occupation number.~\cite{cold_atom,cold_atom1}
The paramagnetic-antiferromagnetic transition
in the spin model can then be mapped to a transition between
a phase with singly occupied sites at small tilt and a density wave phase at large tilt,
which is detectable via occupation measurements using high-resolution imaging or noise correlation measurements.~\cite{cold_atom,noise_cr}
Disorder (spatial inhomogeneity) is naturally present in optical lattices and 
a near-homogeneous system can be realized 
by selecting a smaller sample size;~\cite{cold_atom} it is
then possible to experimentally investigate effects of controlled disorder on quantum phase
transitions in ultracold atomic lattice gases.

%Furthermore, the model is
%experimentally realizable using cold atoms in an optical
%lattice,~\cite{cold_atom} enabling experimental investigations of quantum phase
%transitions and dynamics.

The higher-dimensional random transverse-field Ising antiferromagnets have
similar zero-temperature phases to the one-dimensional case.
Infinite-randomness fixed points accompanied by pronounced quantum Griffiths
singularities also govern the low-temperature physics in higher dimensions with
discrete Ising symmetry.~\cite{Motrunich,Lin_2D,Istvan_2D}
In the presence of
a longitudinal field, the quantum dynamics of rare ordered regions and quantum
Griffiths effects are expected to be suppressed, too.  
In principle, there is
no constraint on the implementation of tree tensor network SDRG in two or more
space dimensions and with different types of symmetry (such as SU(2) symmetry),
as long as disorder effects are relevant.~\cite{Damle,Bilayer,Refael_2D}
Also the MPO technique proposed in this paper for calculating higher moments can be applied in high dimensions.
A potential difficulty in implementing high-dimensional SDRG is to deal with a large number of
block-block couplings resulting from fusion of blocks under the action of the RG.  
Nevertheless, for systems governed by strong disorder, it should be possible to find a criterion
for discarding some block couplings without affecting the low-lying energy
spectrum.

\begin{acknowledgments}
We would like to thank Anders~Sandvik,  Matthias~Vojta, and Frank~Pollmann for useful discussions. 
Y.C.L. is grateful to  Z.-R. Hsu for previous collaboration. This work was supported 
by the Ministry of Science and Technology (MOST) of Taiwan under Grants No. 105-2112-M-002-023-MY3,
105-2112-M-004-002, 104-2112-M-002 -022 -MY3.
We also acknowledge support from the NCTS.

\end{acknowledgments}

\end{document}